# AN EFFICIENT FPGA IMPLEMENTATION OF MRI IMAGE FILTERING AND TUMOUR CHARACTERIZATION USING XILINX SYSTEM GENERATOR

Mrs. S. Allin Christe[1], Mr.M.Vignesh[2], Dr.A.Kandaswamy[3]

[1,2]Department of Electronics & Communication Engineering , PSG College of Technology, Coimbatore,India

*sac@ece.psgtech.ac.in , vigneshhh@gmail.com*

[3] Department of  Biomedical Engineering, PSG College of Technology, Coimbatore,India

*hod@bme.psgtech.ac.in*

## ABSTRACT

*This paper presents an efficient architecture for various image filtering algorithms and tumor characterization using Xilinx System Generator (XSG). This architecture offers an alternative through a graphical user interface that combines MATLAB, Simulink and XSG and explores important aspects concerned to hardware implementation. Performance of this architecture implemented in SPARTAN-3E Starter kit (XC3S500E-FG320) exceeds those of similar or greater resources architectures. The proposed architecture reduces the resources available on target device by 50%.*

## 1. INTRODUCTION

The   handling of digital images has become a subject of widespread interest in different areas such as medical, technological applications and many others. There are lots of examples where image processing helps to analyze, infer and make decisions. The main objective of image processing is to improve the quality of the images for human interpretation, or the perception of the machines independently. This paper focuses on processing an image pixel by pixel and in modification of pixel neighbourhoods and the transformation that can be applied to the whole image or only a partial region. The need to process the image in real time, which is time consuming, leads to this implementation in hardware level, which offers parallelism, and thus significantly reduces the processing time. FPGAs are  increasingly used in modern imaging applications image filtering[1,2], medical  imaging[3,4], image compression[5-7], wireless communication[8,9].The drawback of most of the methods are that they use a high level language for coding. This objective lead to the use of Xilinx System Generator, a tool with a high- level graphical interface under the Matlab, Simulink based blocks which makes it very easy to handle with respect to other software for hardware description [10]. The various applications where image filtering operations applied are noise removal, enhancing edges and contours, blurring and so on. This paper presents an architecture of filtering images for edge detection using System Generator, which is an extension of Simulink and consists of a models called "XILINX

BLOCKS", which are mapped into architectures, entities, signs, ports and attributes, which Scripts file to produce synthesis in FPGAs, HDL simulation and developments tools. The tool retains the hierarchy of Simulink when it is converted into VHDL/Verilog. There are many research works related to image processing and its real time implementation using XSG which uses high end hardware similar to the one used in paper [11] by Sami Hasan, Alex Yakovlev and Said Boussakta *et al,* complicated design used in paper [12] by Zhang Shanshan *et al,* but the proposed design in this work eliminates the design complexity, takes least resource usage and also implemented in low cost basic FPGA device (Spartan 3E).

## 2. XILINX SYSTEM GENERATOR

Xilinx System Generator (XSG) [12,13] is an integrated design Environment (IDE) for FPGAs within the ISE 11.3 development suite, which uses Simulink[14], as a development environment and is presented in the form of model based design. It has an integrated design flow, to move directly to the Bit stream file (*. bit) from Simulink design environment which is necessary for programming the FPGA.

One of the most important features of XSG is that it possesses abstraction arithmetic that is working with representation in fixed point with a precision arbitrary, including quantization and overflow. XSG can only perform simulations as a fixed point double precision type. XSG automatically generates VHDL/Verilog code and a draft of the ISE model being developed. It also generates hierarchical VHDL/Verilog synthesis, floor plan and mapping hardware. In addition to this it also generates a user constraint file (UCF), simulation and testbech and test vectors.

XSG was created primarily to deal with complex Digital Signal Processing (DSP) applications, but it also deals with implementation of many images processing application. The blocks in XSG operate with Boolean values or arbitrary values in fixed point type, for a better approach in hardware implementation. In contrast Simulink works with numbers of double-precision floating point. The connection between XSG blocks and Simulink blocks are the gateway blocks. The Fig.1 shows the broad flow design of XSG.

As previously mentioned, XSG is configured to program the FPGA. The synthesis and implementation of the program are done subsequently. In real time implementation of edge detection on FPGA by Sudeep K C *et al* [15], is done with Spartan3A DSP board but in present work Spartan3E starter kit is used to implement the design with least resource usage. The architecture implemented in this paper is versatile for any edge detection operator unlike the paper [11] by Sami Hasan *et al*, which only deals with Sobel operators. Compared to vehicle image edge detection algorithm hardware implementation on FPGA by Zhang Shanshan *et al* [12], the resource usage by the proposed architechture is reduced by 50%.

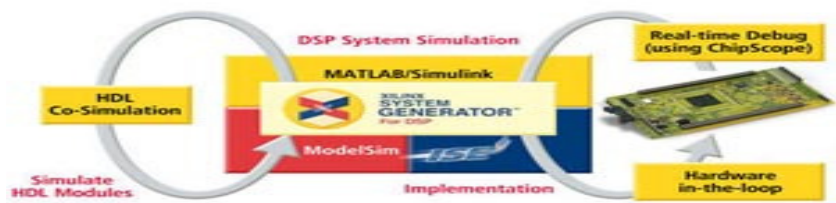

Fig 1: XSG Design Flow

# 3. EDGE DETECTION

Edge detection [16] is one of the most commonly used operations in image analysis, and there are probably more algorithms in literature for enhancing and detecting edges. An edge is point of sharp change in an image, a region where pixel locations have abrupt luminance change i.e. a discontinuity in gray level values. In other words, an edge is the boundary between an object and the background. The shape of edges in images depends on many parameters like the depth discontinuity, surface orientation discontinuity, reflectance discontinuity, illumination discontinuity, and noise level in the images.

The main steps in edge detection are:

1) Filtering which is gradient computation based on intensity values of two points which are susceptible to noise. Filtering reduces noise but there is a trade-off between edge strength and noise reduction.

2) Enhancement is done in order to facilitate the detection of edges, it is essential to determine intensity changes in the neighbourhood of a pixel in an improved manner. Enhancement emphasizes pixels where there is a significant change in local intensity values and is usually performed by computing the gradient magnitude.

3) Detection is done because many points in an image have a nonzero value for the gradient, but not all these points can be considered to be edges. Therefore, some method should be used to determine which points are edge points. Frequently, threshold provides the criterion for detection.

4) Localize/analyze mainly rejects spurious edges include weak but justified edges.
Measuring the relative brightness of pixels in a neighbourhood is mathematically analogous to calculating the derivative of brightness. Brightness values are discrete, not continuous, so we approximate the derivative function. Different edge detection methods use different discrete approximations of the derivative function. The design flow of edge detection using XSG is shown in Fig 2.

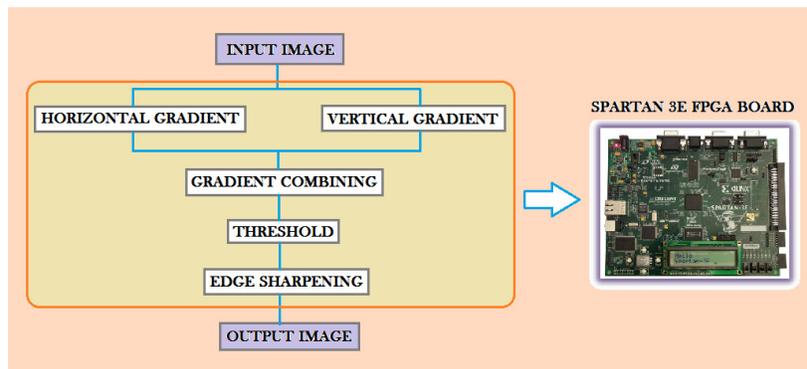

Fig: 2 Design Flow of XSG for image edge detection

The edge detection operators used in this paper is based on the behavioural study of edges with respect to the following two categories:

- Gradient edge detectors (first derivative or classical)
- Zero Crossing or Laplacian (second derivative)

From the first order derivative the Roberts, Prewitt, Sobel and Scharr operators, and from the second order derivative the Laplacian of Gaussian (LoG) operator, also known as the Marr-Hildreth operator and some additional filtering operations has been utilized in this work which is discussed below:

### 3.1. Fist Order Derivative

The Roberts' Cross operator[16] is one of the first edge detectors initially proposed by Lawrence Roberts in 1963 used in image processing and computer vision for edge detection. It a differential operator that approximates the gradient of an image through discrete differentiation which is achieved by computing the sum of the squares of the differences between diagonally adjacent pixels implemented by two 2x2 mask shown in equation (1). These filters have shortest support and more vulnerable to output noise.

$$G_x = \begin{bmatrix} 1 & 0 \\ 0 & -1 \end{bmatrix} \quad G_y = \begin{bmatrix} 0 & 1 \\ -1 & 1 \end{bmatrix} \tag{1}$$

Where $G_x$ is the gradient along x-axis, $G_y$ is the gradient along y-axis. Total Gradient magnitude G and direction can be obtained by using equation (2),

$$G = \sqrt{G_x^2 + G_y^2}, \quad \theta = \tan^{-1}\left(\frac{G_x}{G_y}\right) \tag{2}$$

The Prewitt operator [16] is based on the idea of central difference and is much better than Roberts's operator. It is based on convolving the image with a small, separable, and integer valued filter in horizontal and vertical direction as shown in equation (3). Prewitt's operator has longer support and is less vulnerable to noise.

$$G_x = \begin{bmatrix} -1 & 0 & 1 \\ -1 & 0 & 1 \\ -1 & 0 & 1 \end{bmatrix} G_y = \begin{bmatrix} -1 & -1 & -1 \\ 0 & 0 & 0 \\ 1 & 1 & 1 \end{bmatrix} \tag{3}$$

The Sobel operator is also a central difference with more weights to the central pixels where averaging as given by equation (4). It has an improved noise suppression than Prewitt's operator.

$$G_x = \begin{bmatrix} -1 & -2 & -1 \\ 0 & 0 & 0 \\ 1 & 2 & 1 \end{bmatrix} G_y = \begin{bmatrix} -1 & 0 & 1 \\ -2 & 0 & 2 \\ -1 & 0 & 1 \end{bmatrix} \tag{4}$$

The Sobel operator, while reducing artifacts associated with a pure central differences operator, does not have perfect rotational symmetry.

Scharr operator [17] looked into optimizing this property. Scharr operators' results from an optimization of weighted mean squared angular error in Fourier domain which is done under the condition that resulting filters are numerically consistent. Therefore they really are derivative kernels with symmetry constraints which are shown in equation (5).

$$G_x = \begin{bmatrix} 3 & 10 & 3 \\ 0 & 0 & 0 \\ -3 & -10 & -3 \end{bmatrix} G_y = \begin{bmatrix} 3 & 0 & -3 \\ 10 & 0 & -10 \\ 3 & 0 & -3 \end{bmatrix} \tag{5}$$

## 3.2. Second Order Derivative:

The Laplacian [16] is a 2-D isotropic measure of the second order spatial derivative of an image. The Laplacian of an image highlights regions of rapid intensity change using zero crossing. The LoG is often applied to an image that has first been smoothed with a Gaussian smoothing filter to reduce its sensitivity to noise followed by Laplacian operator. The operator normally takes a single gray level image as input and produces another gray level image as output. The 3x3 kernel approximations to the Laplacian filter is given in equation (6)

$$G = \begin{bmatrix} -1 & -1 & -1 \\ -1 & 8 & -1 \\ -1 & -1 & -1 \end{bmatrix} \tag{6}$$

## 3.3. Additional Filtering Operations

Gaussian blur is usually applied to smoothen the image by reducing the noise in the image. Gaussian 3x3 kernel used is shown in equation (7)

$$G = \begin{bmatrix} 1/16 & 2/16 & 1/16 \\ 2/16 & 4/16 & 2/16 \\ 1/16 & 2/16 & 1/16 \end{bmatrix} \tag{7}$$

Edge sharpening is usually done to strengthen the output image, this leads to connecting the edges to get a sharp output image. 3x3 filter mask applied for edge sharpening is given in equation (8).

$$G = \begin{bmatrix} -1/8 & -1/8 & -1/8 \\ 1/8 & 16/8 & 1/8 \\ -1/8 & -1/8 & -1/8 \end{bmatrix} \tag{8}$$

Thresholding is the simplest method of image segmentation. From a gray scale image, thresholding can be used to create binary images.

# 4. PROPOSED DESIGN

The entire operation of edge detection proposed using Simulink and Xilinx blocks goes through 3 phases,

- Image pre-processing blocks.
- Edge detection using XSG.
- Image post-processing blocks

For the design of filters to meet hardware requirements, it is a must to pre-process the image prior to the main hardware architecture. In the software level simulation using Simulink blocksets alone, where the image is used as a two-dimensional(2D) arrangement such as M x N, there is no need for any image pre-processing, but at hardware level this matrix must be an array of one dimension(1D), namely a vector, where it requires image pre-processing.

## 4.1 Image Pre-processing Blocksets

The model based design used for image pre-processing is shown in Fig.3, the blocks utilized here are discussed below. Input images which could be color or grayscale are provided as input to the File block. A color space conversion block converts RGB to grayscale image and this data which is in 2D is to be converted to 1D for further processing. Frame conversion block sets output signal to frame based data and provided to unbuffer block which converts this frame to scalar samples output at a higher sampling rate.

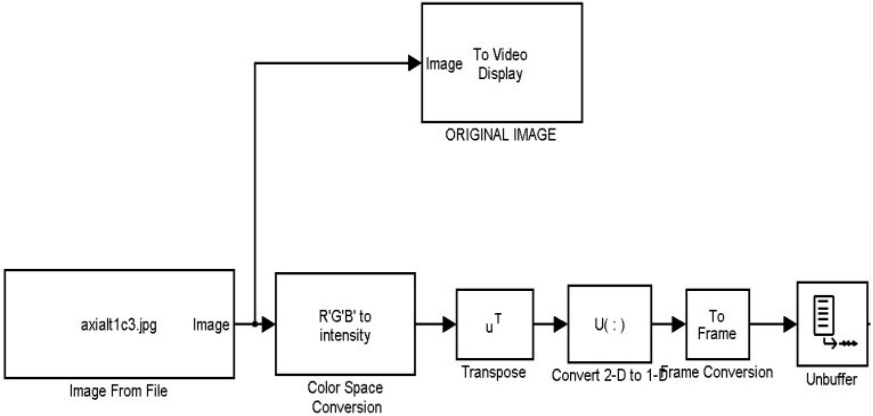

Fig 3: Image Pre-Processing

## 4.2 Edge detection using XSG Blocksets

The model based design [12] using Xilinx blocksets for processing the input image for edge detection is shown in Fig.4,

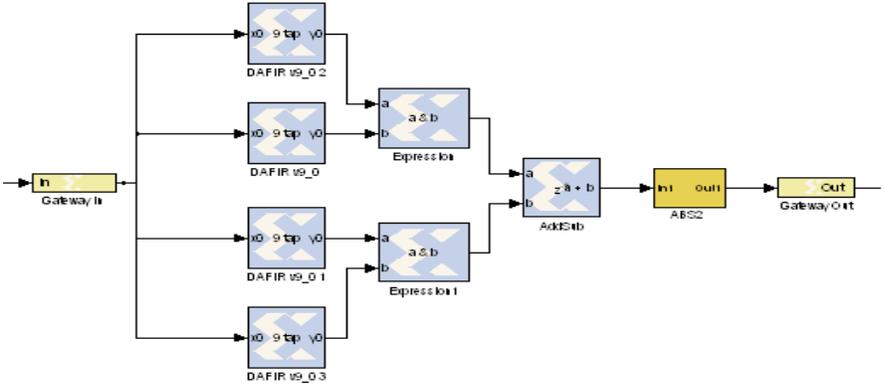

Fig 4: Edge Detection Using XSG

Xilinx fixed point type conversion is made possible by Gateway In block. To perform the edge detection a convolution operation of the input image with a filter mask is to be performed for which a n-tap MAC FIR filter block is used provided with nine programmable coefficients. This is followed by certain arithmetic blocks to merge all the processed data's.

## 4.3 Image post-processing Blocksets

The image post-processing blocks which are used to convert the image output back to floating point type is shown in Fig.5,

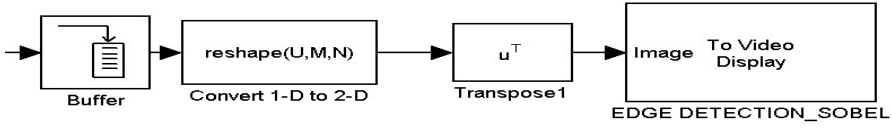

Fig 5: Image Post-Processing

For post-processing it uses a Buffer block which converts scalar samples to frame output at lower sampling rate, followed by a 1D to 2D (matrix) format signal block, finally a sink is used to display the output image back in the monitor, utilizing the Simulink blocksets.

This proposed design architecture has also been utilized in an application oriented design by adding appropriate image post processing blocks as shown in Fig.6 with added features like region of interest (ROI) section which defines the shape and position of ROI and statistical feature extraction for different tissue analysis. The different textural statistics that can differentiate the tissues like mean, variance and standard deviation are computed using equation (9-11)

$$Mean = \mu_{ij} = \frac{\sum_{i=1}^{M} \sum_{j=1}^{N} u_{ij}}{N*N} \qquad (9)$$

$$standard\ deviation = \sigma_{ij} = \frac{\sum_{i=1}^{M} \sum_{j=1}^{N} |u_{ij} - \mu_{ij}|^2}{M*N-1} \qquad (10)$$

$$Variance = (\sigma_{ij})^2 \qquad (11)$$

These parameter are measured for a abnormal region and normal region for 3 different cases of tumours, where M x N is 2D data $u_{ij}$. It could seen from the Table 1 and graphs as shown in Fig 18- Fig 20 there is a clear variation in the properties of 3 cases of tumours considered. This analysis can be extended to further statistical based differentiation among the tumours by considering larger data sets.

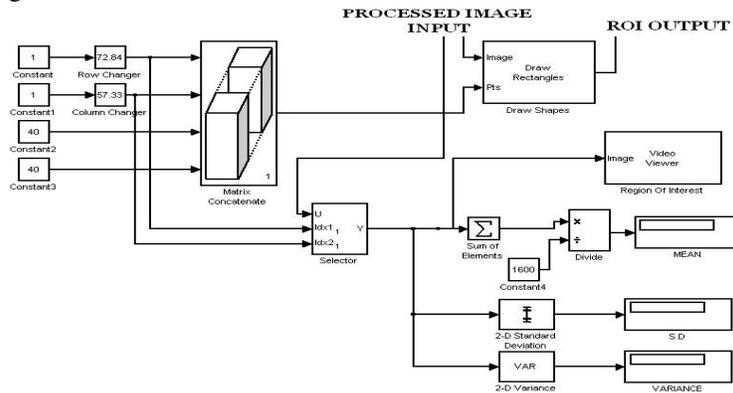

Fig 6: ROI Extraction and Statistical Analysis

## 5. HARDWARE IMPLEMENTATION

The architecture explained above deals only with software simulation level. For implementing this design in a FPGA board the entire module should be converted to FPGA synthesizable one. For that purpose main module for edge detection is converted for JTAG hardware co-simulation, this is done with the help of System generator block which is shown in Fig 7. This block is configured according to the target platform and a bit stream (*.bit) file is generated.

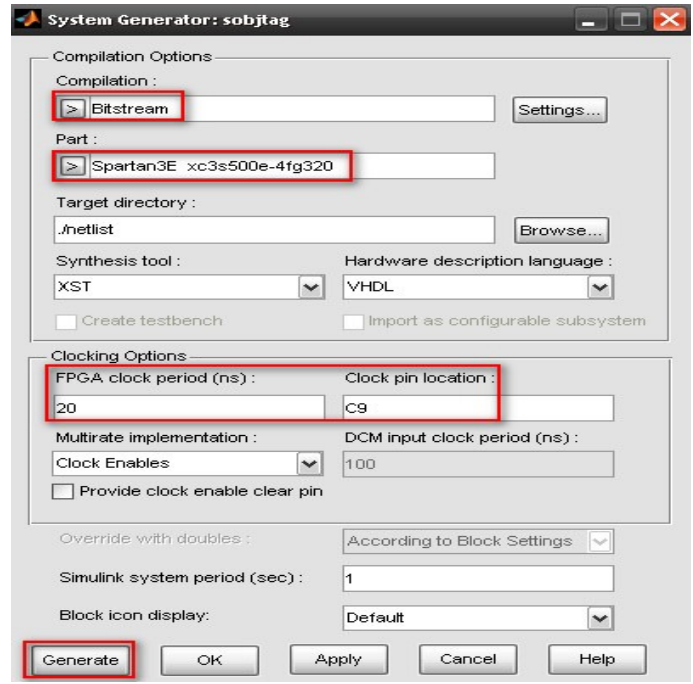

Fig: 7 System Generator Block

After the bit stream file is generated, hardware co-simulation target is selected and in this work, Spartan 3E starter kit (XC3S500E-FG320) is used for board level implementation. The complete design with the edge detection, gaussian blur, thresholding & edge sharpening operations is shown in Fig 8.  The entire architecture with the hardware and software co-simulation design is shown in Fig 9.

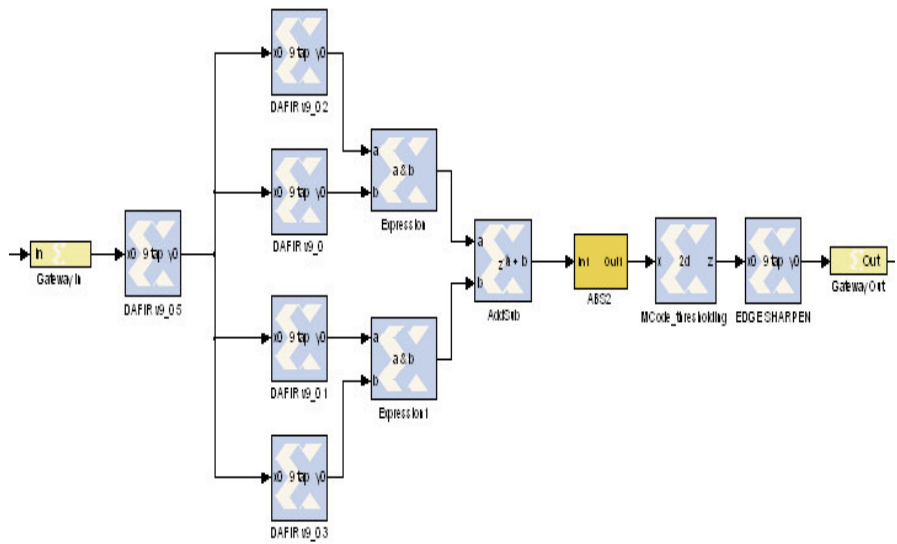

Fig: 8 Complete Design with edge detection, blur, thresholding, & sharpening operation

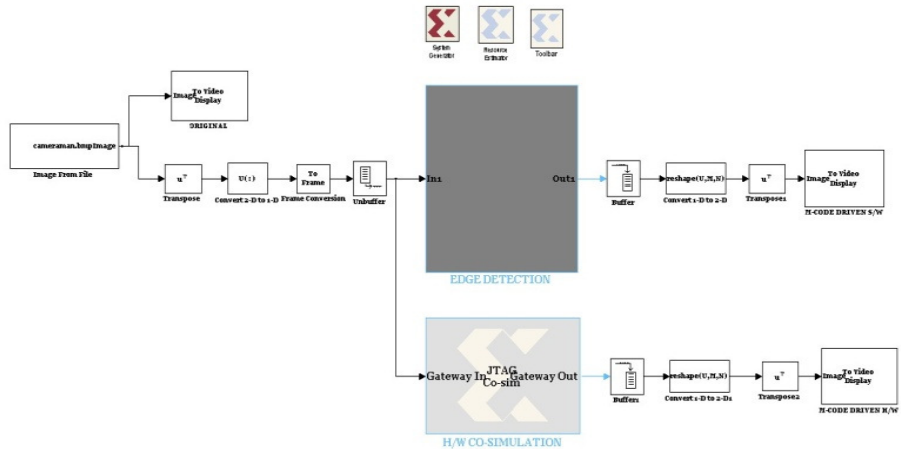

Fig 9: Complete Hardware/Software Co-simulation Design

# 6. RESULTS

The different edge detection operators implemented in this paper are given below along with their corresponding hardware outputs obtained except for Roberts house image which was not clear. The input image utilized for edge detection and outputs of various operators is shown in Fig.10.

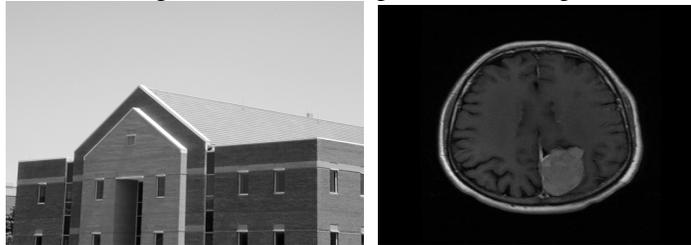

(a): Input Images

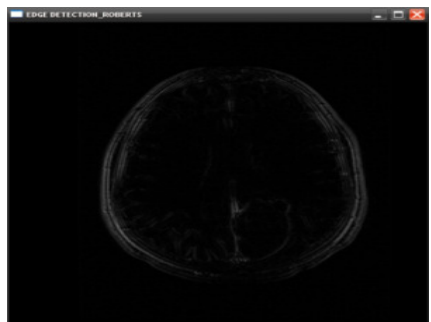

(b) Roberts' Cross Operator Output

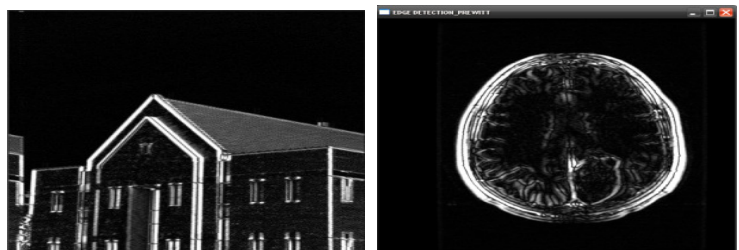

(c) Prewitt Edge Detection Output

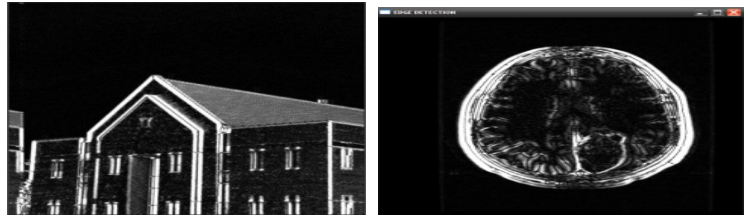

(d) Sobel Edge Detection Output

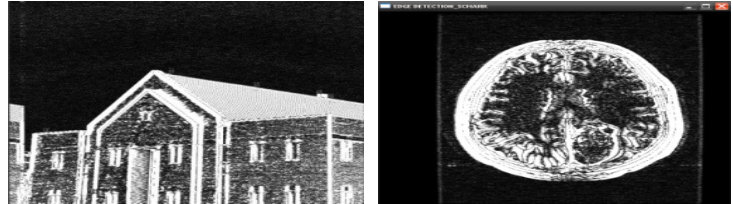

(e) Scharr Operator Output

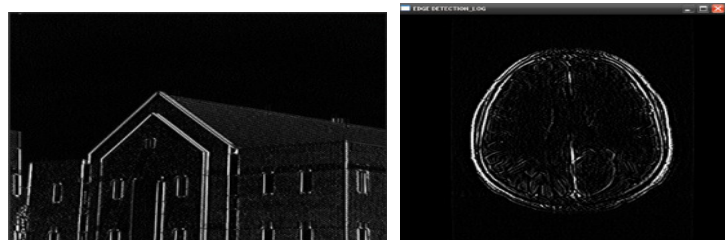

(f) LoG Operator Output

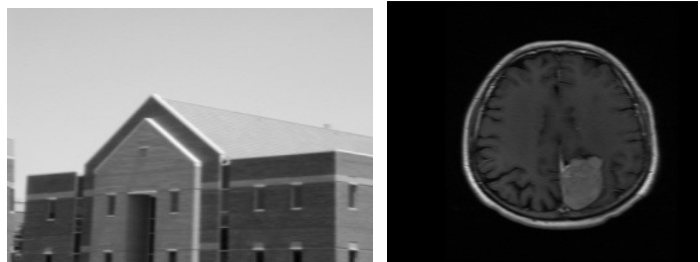

(g) Gaussian Blur Output

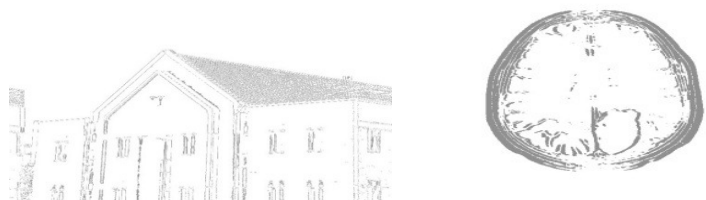

(h) Threshold operation Output

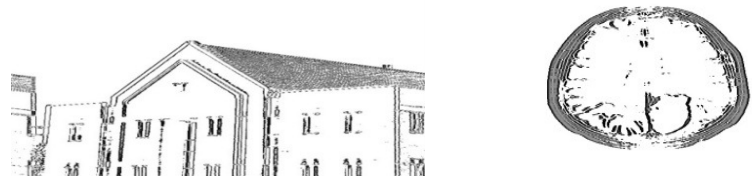

(i) Edge sharpen Output

Fig 10.(a)Input Images of resolution 256 x 256, Edge output using (b)Roberts (c)Prewitt (d)Sobel (e)Scharr (f)LoG (g) Gaussian Blur (h) Threshold (i) Edge Sharpening

The device resource usage is also estimated for this proposed design which is literally reduced by 50% compared vehicle image edge detection algorithm hardware implementation on FPGA by ZhangShanshan and WangXiaohong *et Al* [12].

FPGA Board Selected: XC3S500E-4 FG320
Clock Frequency: 50 MHz

Table 1 Resources utilized

| Resource | Used | Available | Device Usage by proposed design | Device Usage by Zhang [11] |
|---|---|---|---|---|
| Flip Flop | 163 | 9312 | 2 % | 4% |
| Slices | 116 | 4656 | 2.5 % | 5% |
| LUTs | 130 | 9312 | 1.5 % | 3% |
| IOBs | 49 | 232 | 21 % | 16% |

The VHDL code automatically generated by using XSG has got 4547 lines of VHDL coding from this it's clear, that DSP application are more complex and tedious if coded and moreover these DSP application are not logic based, they involve lot of floating/fixed point operation which are hard to be determined. Hence it's clear that by using XSG, program developing & debugging can avoided thereby the design development time is minimized.

The ROI extracted output image is shown in Fig 11, which clearly shows in the MRI-Brain image, with the tumour (ROI) part alone extracted.

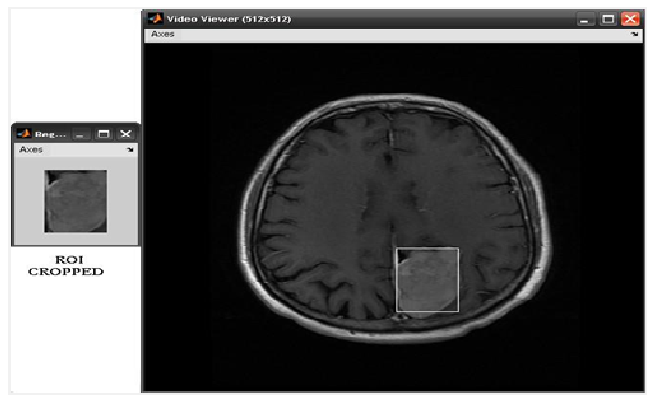

Fig 11: ROI Extracted Output Image

The various textural statistical parameters like mean, variance and standard deviation for different class of tumour (ROI) images are tabulated in Table 2 and its graphical representation is also shown in Fig. 12-14.

Table 2 Textural statistical Parameter estimation of MRI tissues

| case | IMAGE | ROI TYPE | MEAN | VARIANCE | STANDARD DEVIATION |
|---|---|---|---|---|---|
| I | 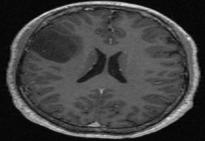 | Normal | $3.18\ e^{-1}$ | $3.08\ e^{-3}$ | $5.55\ e^{-2}$ |
| I | | Abnormal | $2.29\ e^{-1}$ | $4.37\ e^{-3}$ | $6.61\ e^{-2}$ |
| I | 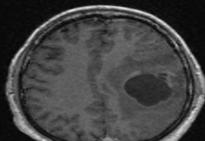 | Normal | $3.23e^{-1}$ | $2.02e^{-3}$ | $4.49e^{-2}$ |
| I | | Abnormal | $2.25e^{-1}$ | $5.17e^{-3}$ | $7.19e^{-2}$ |
| II | 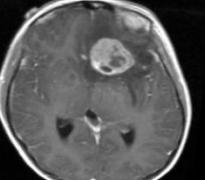 | Normal | $4.10e^{-1}$ | $4.05e^{-3}$ | $6.36e^{-2}$ |
| II | | Abnormal | $5.13e^{-1}$ | $2.48e^{-2}$ | $1.57e^{-1}$ |
| II | 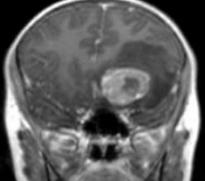 | Normal | $4.20e^{-1}$ | $3.40e^{-3}$ | $5.83e^{-2}$ |
| II | | Abnormal | $5.13e^{-1}$ | $1.60e^{-2}$ | $1.26e^{-1}$ |
| III | 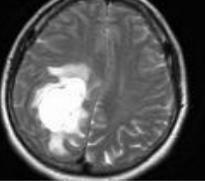 | Normal | $3.91e^{-1}$ | $3.15e^{-3}$ | $5.62e^{-2}$ |
| III | | Abnormal | $8.94e^{-1}$ | $2.80e^{-2}$ | $1.67e^{-1}$ |
| III | 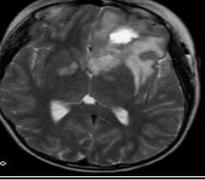 | Normal | $2.38e^{-1}$ | $3.30e^{-3}$ | $5.76e^{-2}$ |
| III | | Abnormal | $6.17e^{-1}$ | $4.90e^{-2}$ | $2.21e^{-1}$ |

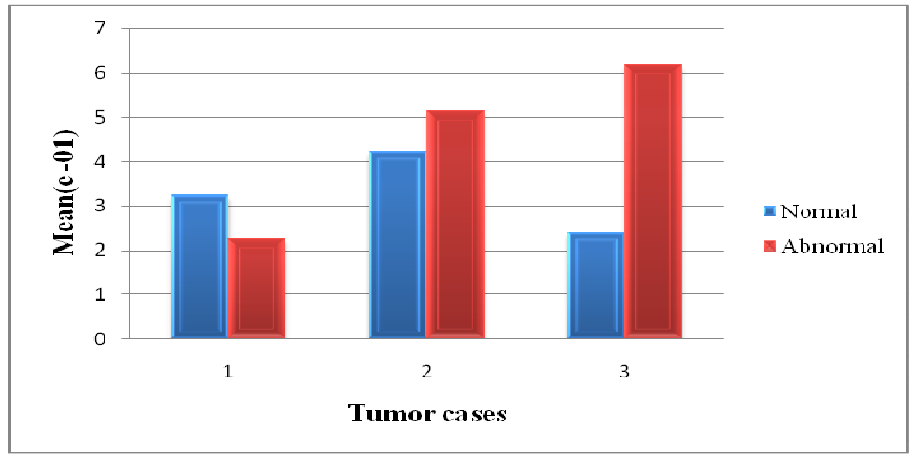

Fig 12. Mean Analysis

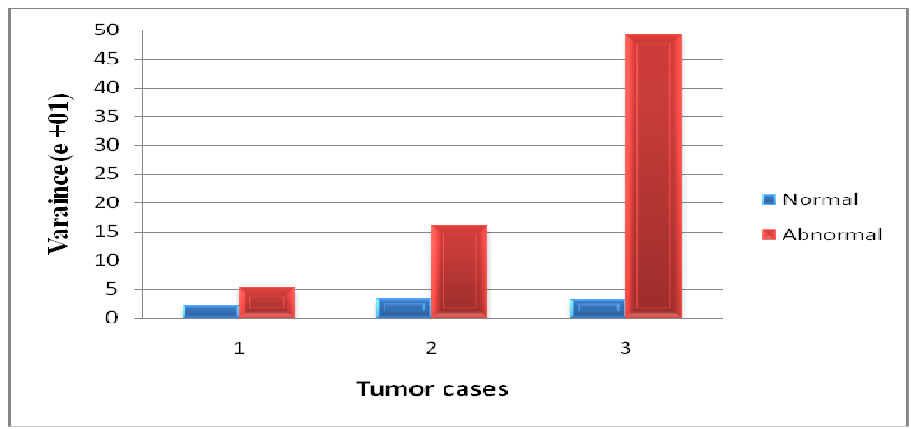

Fig 13. Variance Analysis

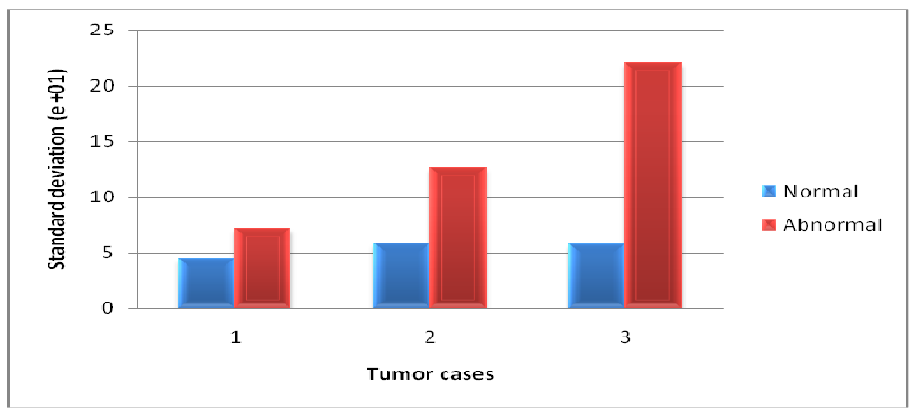

Fig 14. Standard Deviation Analysis

## 7. CONCLUSION

The Xilinx System Generator tool is a new application in image processing, and offers a model based design for processing. The filters are designed by blocks and it even supports Matlab codes through user customizable blocks. It also offers an ease of designing with GUI environment. This tool support software simulation, but most importantly it generates necessary files for implementation in all Xilinx FPGAs, with the parallelism, robust, speed and automatic area minimization. These features are essentials in real time image processing. The design architecture used in this paper can be used for all Xilinx FPGA Kit with proper user configuration in System generator block and could be extended to real time image processing.

## Authors

**S.Allin Christe** is working as an Assistant Professor (S.Gr) in the Department of Electronics   & Communication Engineering of PSG College of Technology, Coimbatore, India. She is pursuing her Ph.D degree from Anna University, Chennai, India. Her research interests include Image Processing, VLSI Design and Soft computing.

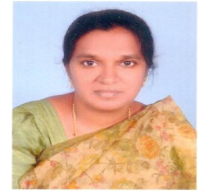

**Vignesh M** graduated from PSG College of    Technology in 2011 from Dept. of Electronics and Communication. His areas of interests include Sytem On Chip, ASIC Design and Image Processing. His research areas include implementation of image processing algorithms on Embedded and Reconfigurable [FPGA]  hardware.He was a post graduate student in Dept. of Electronics and Communication in  PSG College of Technology between 2009-2011.

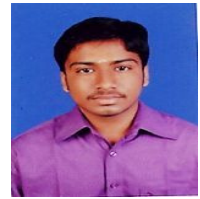

**Dr.Kandaswamy Arumugam** is Professor and Head , Department of Biomedical Engineering, PSG College of Technology, Coimbatore, India. He has 38 years of teaching experience. He has published more than 80 papers in national and international journals and conference proceedings. His fields of interest are Image Processing Applications in Medicine and wireless communication systems.

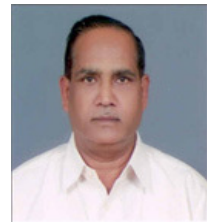